# Enhanced Logic Performance with Semiconducting Bilayer Graphene Channels

Song-Lin Li,[a] Hisao Miyazaki,[a,b] Hidefumi Hiura,[a,c] Chuan Liu,[a] and Kazuhito Tsukagoshi [a,b,1]

[a] International Center for Materials Nanoarchitectonics (MANA), National Institute for Materials Science, Tsukuba, Ibaraki 305-0044, Japan

[b] CREST, Japan Science and Technology Agency, Kawaguchi, Saitama 332-0012, Japan

[c] Green Innovation Research laboratories, NEC Corporation, Tsukuba, Ibaraki 305-8501, Japan

E-mail: li.songlin@nims.go.jp or tsukagoshi.kazuhito@nims.go.jp

ABSTRACT. Realization of logic circuits in graphene with an energy gap (EG) remains one of the main challenges for graphene electronics. We found that large transport EGs (>100 meV) can be fulfilled in dual-gated bilayer graphene underneath a simple alumina passivation top gate stack, which directly contacts the graphene channels without an inserted buffer layer. With the presence of EGs, the electrical properties of the graphene transistors are significantly enhanced, as manifested by enhanced on/off current ratio, subthreshold slope and current saturation. For the first time, complementary-like semiconducting logic graphene inverters are demonstrated that show a large improvement over their metallic counterparts. This result may open the way for logic applications of gap-engineered graphene.

---

[1] To whom correspondence should be addressed. E-mail: li.songlin@nims.go.jp or tsukagoshi.kazuhito@nims.go.jp

Since the isolation of graphene in 2004,[1] its suitability as a channel material in the post-silicon era has been an interesting but challenging topic.[2-4] Besides ultra-high carrier mobility,[5,6] graphene also has advantages over other candidates (*e.g.*, carbon nanotubes) for nanoelectronics, such as being free of chirality and diameter control, and availability of large-area sheets.[7] However, the intrinsic metal nature and lack of energy gaps (EGs)[8] make it difficult to turn off a graphene field-effect transistor (FET) and to incorporate it into logic electronic devices. Thus, logic applications with graphene, especially with a substantial EG, are rarely explored,[9-11] despite the rapid progress in radio-frequency analog applications in recent years.[12,13] It is necessary to introduce an EG into graphene to reduce the off-state current. For this purpose, two main schemes have been proposed: lateral confinement[14-16] and inversion symmetry breaking.[17-20] In the former scheme, sizable EGs were successfully opened by cutting the graphene into ultra-narrow ribbons (GNRs), and were demonstrated by high-performance GNR-FETs.[21,22] However, this also drives the channel width into a lithography inaccessible range (<10 nm) and causes large carrier-mobility degradation. In contrast, no strict dimensional limitation is imposed with the latter scheme. For instance, within perpendicular electric fields the band structure would vary considerably in bilayer graphene (BLG) sheets, which would lead to a gap between the conduction and valence bands.[23,24]

A complementary logic inverter, also known as a NOT gate, is one of the building block of integrated circuits. It consists of two FETs with opposite carrier types and is usually used as a prototype to test emerging electronic materials.[25-27] In a significant advancement,[10] Traversi and coworkers first demonstrated graphene inverters with complementary-like geometry. However, only a metallic single layer graphene (SLG) channel was used. In another work,[11] although a dual-gated BLG structure was employed, no substantial EG was observed. These initial studies also had a common limitation, *i.e.*, using thick dielectric coupling layers (300 nm $SiO_2$ in Ref. 10 and 50 nm AlOx in Ref. 11), which normally requires high operation bias (input voltage, $V_{IN}$), and thus none were able to reach device performance with >0.05 voltage gain and >20% output voltage ($V_{OUT}$) swing. Three factors are essential for an inverter unit: the switching characteristics, gating efficiency and circuitry geometry. Recently, we have been able to significantly enhance the inverter performance (*e.g.*, voltage gain of 6) in gapless SLG devices by adopting a highly efficient alumina dielectric.[28] It is anticipated that the performance would be further improved if a sizable EG could be introduced into the devices.

In this work, we report the electric properties of the first semiconducting graphene logic inverters based on a buffer-free top gate dielectric method. Sufficiently strong perpendicular electric fields can be obtained under low top gate (TG) voltage ($V_{TG}$<1.5 V), due to the high capacitive efficiency of TG stack. We find that the alumina dielectric can effectively suppress the extrinsic charges and a >100 meV transport EG is obtained, comparable to the state-of-art semiconducting BLG FETs,[29] in which a buffer layer, such as polymer NFC, is required to separate graphene from oxide dielectric. In contrast to the metallic SLG inverters that exhibit high performance at high supply bias ($V_{DD}$), the BLG devices exhibit superior electrical characteristics in the low $V_{DD}$ region with a maximum $V_{OUT}$ swing of 80% and a highest voltage gain of 7 . The existence of the EG can greatly reduce the operating current, which promises to solve the issue of power consumption confronted by SLG inverters. Most importantly, the suitable operating $V_{DD}$ level is around 0.5 to 1 V, which indicates high compatibility with the present microchip CMOS bias level and that planned for the future.[2]

RESULTS AND DISCUSSION

Although the existence of an EG in perpendicular electric field biased BLGs has been predicted by theoretical studies (Fig. 1d)[17-19] and confirmed by optical experiments,[23] the expected insulating state has been elusive in transport measurements.[24] In principle, the off-state current is closely correlated to the residual carrier concentration $n_0$, and should ideally approach zero at the charge neutrality point (CNP) in the biased BLGs, due to disappearance of density of states.[30] However, in real devices, large charge inhomogeneity (cause electron and hole puddles at CNP) is present due to extrinsic charge sources (such as interfacial trapped charges, residing charges inside substrates, and charged adsorbates). Such scattering sources can also degrade mobility.[31,32] In this sense, to exclude the extrinsic charge sources is important not only to reduce $n_0$, but also to improve mobility $\mu$. By inserting a buffer layer (polymer NFC) between the high-κ dielectric and graphene, a low $n_0$ of $2.8×10^{11}$/cm$^2$ and a high room temperature (RT) $μ$ of 7300 cm$^2$/Vs have been achieved in SLG.[33] Application of the same technique to BLG resulted in an observed EG of >130 meV.[29]

In our experiment, a thin and dense natural alumina passivation layer was adopted as the TG dielectric, not only for the purpose of high gating efficiency,[28,34] but also to preserve the intrinsic behavior of graphene. It was reported that a lower $n_0$ of $2.3×10^{11}$/cm$^2$ and a higher RT $μ$ (8600 cm$^2$/Vs) are observed in SLGs under a direct contact medium-κ (4.5–8.9) alumina dielectric,[35] which implies a simpler buffer-free solution to suppress the extrinsic charge sources. We identified that even without additional buffer layers, a large EG could be also achieved in our BLG samples exfoliated from natural graphite and covered with the natural alumina TG dielectric. High-purity natural graphite may contain fewer defects, and the formation of alumina dielectric by natural oxidation[34,35] may introduce fewer charge impurities compared with conventional methods. The medium permittivity of alumina leads to reduced interfacial polar phonons, which may also help to maintain the intrinsic $μ$. All factors are critical to the preservation of intrinsic behavior and the realization of a large transport EG.

The effect of the buffer-free alumina passivation dielectric on the graphene channels was first verified on SLG flakes. Without deducting the series resistance ($R_s$, from the TG uncovered graphene regions near source and drain) and the contact resistance ($R_c$), the maximum on/off ratio (at bottom gate voltage, $V_{BG}$, =-30 V) can reach 8 and 14 at RT and 77 K, respectively (Figures 1e and 1f). The $R$ curves exhibits rather sharp peaks, which indicates small residual charge densities at the CNP. With consideration of the quantum capacitance, the $R$ curves were fitted using the expression[35]

$$R_{tot} = R_c + R_s + \frac{L}{We\mu\sqrt{n_0^2 + n^2}} \quad (1),$$

where $R_{tot}$ is the total resistance, $L$ is the channel length, $W$ is the channel width, and $n$ is the gate-induced carrier density. $n_0$ and $μ$ at $V_{BG}$=0 were estimated to be $3.5×10^{11}$/cm$^2$ and 5000 cm$^2$/Vs under ambient conditions (292 k, air) and were improved to $1.4×10^{11}$/cm$^2$ and 8500 cm$^2$/Vs in liquid nitrogen (77 K), which suggests that the alumina passivation dielectric has excellent inertness toward the graphene channel, bringing a small number of extrinsic charges. Therefore, in addition to the high coupling efficiency and the simplicity in simultaneously fabricating both the TG and dielectric,[28,34] this TG stack technique also provides a buffer-free operation in preserving the intrinsic behavior of the graphene channels.

The effect of the novel dielectric was then verified on BLG flakes. Figures 1g and 1h show the $R$ characteristics as a function of $V_{TG}$ (from -1.5 to 1.5 V) under different BG voltages ($V_{BG}$=-30, 0, and 30 V). Contrary to the SLG FETs in which the $R$ peak ($R_{CNP}$) slightly decreases with $V_{BG}$ (because $R_s$ decreases when increasing $V_{BG}$), the $R_{CNP}$ of the BLG FETs increased considerably with $V_{BG}$ (more strictly, perpendicular displacement field $D_z$), irrespective of the measurement temperature. Such a tendency is consistent with the theoretical prediction of opening and expansion of EG with $D_z$.[17-19] At RT, $R_{CNP}$ increases 7.7 times when $V_{BG}$ is changed from 0 ($V_{TG}$~0, $D_z$~0) to -30 V ($V_{TG}$=1.3 V, $D_z$=1.3 V/nm). Assuming that the off current is dominated by the thermionic emission process through a Schottky barrier, as suggested by Ref. 29, a barrier variation around 52 meV can be extracted accordingly. Typically, the barrier has a maximum when the Fermi level of the contact is aligned to the middle of EG of the channel (*i.e.*, EG is twice that of the barrier). This indicates that the transport EG at $D_z$=1.3 V/nm is at least 100 meV, which is comparable to the value obtained in Ref. 29, but realized under lower $V_{TG}$ bias and a much simpler TG fabrication process. We note that the observed transport gap is still much smaller than that revealed by optical measurements (*ca.* 130 meV) under the same $D_z$ condition, and better FET performance would be expected if the extrinsic factors in BLG could be further minimized. It should also be kept in mind that although the observed value is large for biased BLG, it is relatively small compared with conventional semiconductor materials. The estimated 52 meV barrier is also rather low with respect to the RT thermal activation energy $k_BT$~26 meV and therefore is insufficient to effectively block thermal carrier injection at RT. Therefore, the EG obtained here cannot support high performance RT operation (Supporting Information for RT characteristics). In order to demonstrate the effect of the small EG in graphene, the following measurements were all performed in liquid nitrogen at 77 k to reduce the thermal activation energy. The nitrogen atmosphere also enables suppression of the gating hysteresis caused by oxygen and humidity desorption/absorption. Upon replacing SLG with BLG for the channel, better on/off ratios from 14 to 400 and enhanced subthreshold slopes from 600 to 160 mV/decade were obtained.

Opening of an EG in the BLG FETs is also reflected by the different output characteristics from the SLG FETs. Figure 2 displays a series plot of $I_{DS}$-$V_{DS}$ curves for the both types of FETs. For the gapless SLG FETs, the unique current saturation behavior and second linear region[36] is clearly present in small $\Delta V_{TG} = \left| V_{TG} - V_{TG}^{0} \right|$ curves in Fig. 2a, indicating the absence of barrier for minority carrier injection[3,36]. The $I_{DS}$ curves also cross with one another when $V_{DS}$ is sufficiently large, resulting in a zero or even negative transconductance, which implies the loss of effective control of gate on channel current. In contrast, the current saturation characteristic is largely improved in the BLG FET with the presence of EG, as reflected by the smaller slope and more extended current saturation region in Fig. 2b. At $V_{TG} - V_{TG}^{0}$ = -0.7 V and $V_{DS}$= -1 V (indicated by solid dots), the drain conductance $g_{DS} = (\partial I_{DS} / \partial V_{DS})|_{V_{TG}, V_{BG}}$ values for SLG and BLG FETs are 620 and 31 μA/V, respectively, indicating a 20 times improvement of current saturation. The improved current saturation behavior may also lead to a better intrinsic gain for radio-frequency analog applications.[3]

Complementary-like semiconducting logic inverters were fabricated based on the ambipolarity of graphene FETs,[28] as pictured in Fig. 3a and with the operating principle shown in Figs. 3b–3c. The common TG and source are used as input and output electrodes, respectively. For comparison, the electrical performance characteristics of both the metallic SLG and semiconducting BLG inverters are shown in Figs. 3d–3i. An apparent improvement in $V_{OUT}$ swing is observed for the BLG device in the measured $V_{DD}$ ranges due to suppressed $\sigma_0$ (Figs. 3d and 3g), and this trend is summarized in Figure 3j. When $V_{DD}$ is increased from 0 to 2 V, the $V_{OUT}$ swing (defined as $(V_{OUT}^{max} - V_{OUT}^{min})/V_{DD} \times 100\%$) increases slightly from 30 to 45% for the SLG device, while for the BLG device it has a maximum at 80% and then decreases to 55% as $V_{DD}$ increases until 2 V. A sharper voltage inversion (especially in the low $V_{DD}$ region) is observed when BLG channels are used, which becomes more apparent when comparing the voltage gain (defined as the maximum $-dV_{OUT}/dV_{IN}$) in Figs. 3e and 3h. Specifically, the voltage gain increases by factor of six (from 1 to 6) at $V_{DD}$=0.5 V in the BLG device, which suggests that the role of the EG is significant. The voltage gain for both devices also has a different dependence on $V_{DD}$ (Fig. 3k). It is approximately linear for the SLG device, while becomes non-monotonic with a maximum of ~7 at $V_{DD}$=1 V for the BLG device. Furthermore, the operating current also decreases with the introduction of the EG. The current minimum is reduced by one order of magnitude at $V_{DD}$=0.5 V (Fig. 3l). Overall, the BLG inverter exhibits an obvious improvement of inverter performance in the low $V_{DD}$ region with the presence of the EG. In the high $V_{DD}$ region, the performance of the BLG device degrades to that of the SLG device.

We now qualitatively discuss the factors that determine the device performance. If $R_s$ and $R_c$ are neglected and a linear $\sigma - V_{TG}$ relation is assumed ($\sigma = \sigma_0 + k|V_{TG} - V_{TG}^0|$, where $k$ is the tunability of conductance through gate voltage, $\sigma_0$ is the off-state conductance, and $V_{TG}^0$ is the CNP position), then the voltage gain and $V_{OUT}$ swing can be simply expressed as (See supporting information for derivation)

$$V_{out} \text{ Swing} = \frac{1}{1 + 2\sigma_0 / k\Delta} \quad (2),$$

$$\text{Voltage Gain} = \frac{V_{DD}}{2\sigma_0 / k + \Delta} \quad (3),$$

where $\Delta$ is the CNP splitting between the two FETs along the $V_{IN}$ axis. The improvement of $V_{OUT}$ swing and voltage gain in the BLG devices in low $V_{DD}$ regions relies on two factors; 1) high capacitive efficiency (*ca.* 920 nF/cm$^2$) which enables a high $k$ value, and 2) introduction of EG which leads to a small $\sigma_0$ (or large $R_{CNP}$). At high $V_{DD}$, both the $\sigma_0$ and $k$ decrease where extra carriers are induced at the off state,[3,36] leading to a performance degradation, as shown in Figs. 3j and 3k. Interestingly, a large $\Delta$ is beneficial for $V_{OUT}$ swing but unfavorable for voltage gain, and at fixed $V_{DD}$ there is a tradeoff between them.

To demonstrate the logic operations on the BLG device, dynamic response was also tested under 0.1–10 kHz square-waveform stimulus (Figure 4). Here, $V_{DD}$ was set at 1 V and $V_{BG}$ was selected at -6 V to match $V_{IN}$ and $V_{OUT}$. Such a biasing condition leads to high and low logic voltage levels at 0.8 and 0.3 V, respectively. The ability of exhibiting matched input and output signals under a decreased $V_{DD}$ level further reduces the power consumption with respect to previous SLG device.[28] At frequencies of 0.1 and 1 kHz, the output response exhibits well-defined inversion behavior of the input signal, which indicates its excellent Boolean operation capability. At higher frequencies (*e.g.* 10 kHz), small output deformation is observed that can be ascribed to the large stray capacitance of the measurement system.

In this work, the introduction of the EG into the BLG channel significantly improves the low-bias performance, with the appropriate $V_{DD}$ region around 0.5–1 V. In particular, at $V_{DD}$= 0.5 V, the gain and $V_{OUT}$ swing reach *ca.* 6 and 80%, respectively. It is important to note that the present semiconductor microchip $V_{DD}$ criterion is 1 V and is expected to be reduced to 0.7–0.8 V in 2024 with further feature size scaling down.[2] Therefore, semiconducting graphene can satisfy the low bias requirement for future microchips. Significantly, the atomic channel thickness of graphene greatly reduces the effective vertical depletion length and helps to minimize the characteristic scale length, which may allow graphene FETs to be scaled to shorter channel lengths and higher speeds without encountering the adverse short channel effects.[3,37] In this regard, graphene may be an exclusively suitable channel material with respect to dimensionality for the post-silicon era. The present work also signifies a relatively simple fabrication technique for logic components, because it requires neither a doping process to realize the complementary geometry, nor rigorous dimensional control to create the EGs. Most importantly, it is a real lithography-compatible technique, making it superior to those integrated with nanowires or nanotubes in which the location and alignment control remain an issue. The rapid development of large-area graphene growth techniques in recent years, especially with chemical vapor deposition,[7] has further paved the way for electronic applications employing graphene. Moreover, several recent experiments, such as chemical modification[38,39] and structure perforation (graphene nanomesh)[40,41] have also showed promising results with respect to the creation of large EGs in graphene. For instance, graphene can be transformed from a highly conductive semimetal into an insulator by reversible hydrogenation.[38] A high on/off ratio of 2 orders of magnitude is observed in graphene nanomesh samples at RT.[40] These novel methods, in principle being lithography compatible, would also lend themselves to gap-engineered graphene electronics, if reasonably high carrier mobility could be retained as the EGs are opened.

CONCLUSIONS

We have demonstrated a simple and buffer-free TG dielectric technique, which is able to preserve the intrinsic nature and achieve a large transport EG in BLG. Based on the EG, the electrical characteristics of the complementary-like semiconducting bilayer graphene logic inverters were reported for the first time. Although the EG is not sufficiently large to support high-performance RT operation, the low T results, which do show significant improvement for low-bias performance over metallic inverters, still provide a strong indication of the possibility of gapped graphene as an excellent channel material for future integrated circuits. With further advances in gap engineering, the graphene device may find applications in logic circuits.

METHODS

The graphene flakes used in this experiment were exfoliated from natural graphite. Conventional electron beam lithography and thermal evaporation (50 nm Au / 5 nm Ti) were used to define the interconnection leads on devices.[28,34,42] The structure of the dual-gated FETs (Fig. 1a) resemble those in previous reports,[24,36] in which a graphene sheet is sandwiched by a TG and a degenerately doped silicon bottom gate (BG, with 90-nm $SiO_2$ dielectric). Figures 1b and 1c show optical microscope images for a typical device before and after electrode formation. An array of five FET channels (FET1–FET5) were simultaneously defined with equal dimensions on a single graphene sheet. The experimental channel widths and lengths ranged from 250–350 nm and 1.2–3.6 μm, respectively. The dimension between devices may vary slightly and adapt to the sizes of exfoliated graphene flakes.

*Acknowledgement*: This work was supported in part by a Grant-in-Aid for Scientific Research (No. 21241038) from the Ministry of Education, Culture, Sports, Science and Technology (MEXT) of Japan, and by the Funding Program for World-Leading Innovative R&D on Science and Technology (FIRST Program) from the Japan Society for the Promotion of Science (JSPS).*Supporting Information Available*: Reliability of top gate, detailed electrical characteristics for SLG and BLG FETs, mobility fittings with quantum capacitance, performance for BLG inverters under ambient environment, and derivation of qualitative expressions for $V_{OUT}$ swing and voltage gain.

REFERENCES AND NOTES

1. Novoselov, K.; Geim, A.; Morozov, S.; Jiang, D.; Zhang, Y.; Dubonos, S.; Grigorieva, I.; Firsov, A. Electric Field Effect in Atomically Thin Carbon Films. *Science* 2004, *306*, 666–669.
2. The International Technology Roadmap for Semiconductors. http://www.itrs.net/Links/2009ITRS/Home2009.htm, (Semiconductor Industry Association, 2009).
3. Schwierz, F. Graphene Transistors. *Nat. Nanotechnol.* 2010, *5*, 487–496.
4. Fuhrer, M. S.; Lau, C. N.; MacDonald, A. H. Graphene: Materially Better Carbon. *MRS Bull.* 2010, *35*, 289–295.
5. Du, X.; Skachko, I.; Duerr, F.; Luican, A.; Andrei, E. Y. Fractional Quantum Hall Effect and Insulating Phase of Dirac Electrons in Graphene. *Nature* 2009, *462*, 192–195.
6. Bolotin, K. I.; Sikes, K. J.; Hone, J.; Stormer, H. L.; Kim, P. Temperature-Dependent Transport in Suspended Graphene. *Phys. Rev. Lett.* 2008, *101*, 096802.
7. Bae, S.; Kim, H.; Lee, Y.; Xu, X.; Park, J.-S.; Zheng, Y.; Balakrishnan, J.; Lei, T.; Kim, H. R.; Song, Y. I. *et al.* Roll-to-Roll Production of 30-Inch Graphene Films for Transparent Electrodes. *Nat. Nanotechnol.* 2010, *5*, 574–578.
8. Ando, T. Physics of Graphene. *Prog. Theor. Phys. Suppl.* 2008, *176*, 203–226.
9. Sordan, R.; Traversi, F.; Russo, V. Logic Gates with a Single Graphene Transistor. *Appl. Phys. Lett.* 2009, *94*, 073305.
10. Traversi, F.; Russo, V.; Sordan, R. Integrated Complementary Graphene Inverter. *Appl. Phys. Lett.* 2009, *94*, 223312.
11. Harada, N.; Yagi, K.; Sato, S.; Yokoyama, N. A Polarity-Controllable Graphene Inverter. *Appl. Phys. Lett.* 2010, *96*, 012102.
12. Meric, I.; Baklitskaya, N.; Kim, P.; Shepard, K. RF Performance of Top-Gated, Zero-Bandgap Graphene Field-Effect Transistors. *IEEE International Electron Devices Meeting,* 2008.
13. Lin, Y.-M.; Dimitrakopoulos, C.; Jenkins, K. A.; Farmer, D. B.; Chiu, H.-Y.; Grill, A.; Avouris, P. 100-GHz Transistors from Wafer-Scale Epitaxial Graphene. *Science* 2010, *327*, 662.
14. Son, Y.-W.; Cohen, M. L.; Louie, S. G. Energy Gaps in Graphene Nanoribbons. *Phys. Rev. Lett.* 2006, *97*, 216803.
15. Barone, V.; Hod, O.; Scuseria, G. E. Electronic Structure and Stability of Semiconducting Graphene Nanoribbons. *Nano Lett.* 2006, *6*, 2748–2754.
16. Yang, L.; Park, C.-H.; Son, Y.-W.; Cohen, M. L.; Louie, S. G. Quasiparticle Energies and Band Gaps in Graphene Nanoribbons. *Phys. Rev. Lett.* 2007, *99*, 186801.
17. McCann, E. Asymmetry Gap in the Electronic Band Structure of Bilayer Graphene. *Phys. Rev. B* 2006, *74*, 161403.
18. Castro, E. V.; Novoselov, K. S.; Morozov, S. V.; Peres, N. M. R.; Dos Santos, J. M. B. L.; Nilsson, J.; Guinea, F.; Geim, A. K.; Neto, A. H. C. Biased Bilayer Graphene: Semiconductor with a Gap Tunable By the Electric Field Effect. *Phys. Rev. Lett.* 2007, *99*, 216802.
19. Min, H.; Sahu, B.; Banerjee, S. K.; MacDonald, A. H. *ab initio* Theory of Gate Induced Gaps in Graphene Bilayers. *Phys. Rev. B* 2007, *75*, 155115.
20. Giovannetti, G.; Khomyakov, P. A.; Brocks, G.; Kelly, P. J.; van den Brink, J. Substrate-Induced Band Gap in Graphene on Hexagonal Boron Nitride: *ab initio* Density Functional Calculations. *Phys. Rev. B* 2007, *76*, 073103.
21. Li, X.; Wang, X.; Zhang, L.; Lee, S.; Dai, H. Chemically Derived, Ultrasmooth Graphene Nanoribbon Semiconductors. *Science* 2008, *319*, 1229–1232.
22. Wang, X.;Ouyang, Y.; Li, X.; Wang, H.; Guo, J.; Dai, H. Room-Temperature All-Semiconducting sub-10-nm Graphene Nanoribbon Field-Effect Transistors. *Phys. Rev. Lett.* 2008, *100*, 206803.
23. Zhang, Y.; Tang, T.-T.; Girit, C.; Hao, Z.; Martin, M. C.; Zettl, A.; Crommie, M. F.; Shen, Y. R.; Wang, F. Direct

Observation of a Widely Tunable Bandgap in Bilayer Graphene. *Nature* 2009, *459*, 820–823.
24. Oostinga, J. B.; Heersche, H. B.; Liu, X.; Morpurgo, A. F.; Vandersypen, L. M. K. Gate-Induced Insulating State in Bilayer Graphene Devices. *Nat. Mater.* 2008, *7*, 151–157.
25. Derycke, V.; Martel, R.; Appenzeller, J.; Avouris, P. Carbon Nanotube Inter- and Intra-Molecular Logic Gates. *Nano Lett.* 2001, *1*, 453–456.
26. Nam, S.; Jiang, X.; Xiong, Q.; Ham, D.; Lieber, C. M. Vertically Integrated, Three-Dimensional Nanowire Complementary Metal-Oxide-Semiconductor Circuits. *Proc. Natl. Acad. Sci. U. S. A.* 2009, *106*, 21035–21038.
27. Singh, T.B.; Senkarabacak, P.; Sariciftci, N. S.; Tanda, A.; Lackner, C.; Hagelauer, R.; Horowitz, G. Organic Inverter Circuits Employing Ambipolar Pentacene field-Effect Transistors. *Appl. Phys. Lett.* 2006, *89*, 033512.
28. Li, S.-L.; Miyazaki, H.; Kumatani, A.; Kanda, A.; Tsukagoshi, K. Low Operating Bias and Matched Input–Output Characteristics in Graphene Logic Inverters. *Nano Lett.* 2010, *10*, 2357–2362.
29. Xia, F.; Farmer, D. B.; Lin, Y.-M.; Avouris, P. Graphene Field-Effect Transistors with High On/Off Current Ratio and Large Transport Band Gap at Room Temperature. *Nano Lett.* 2010, *10*, 715–718.
30. Koshino, M. Electronic transport in bilayer graphene. *New J. Phys.* 2009, *11*, 095010.
31. Hwang, E. H.; Das Sarma, S. Single-Particle Relaxation Time versus Transport Scattering Time in a Two-Dimensional Graphene Layer. *Phys. Rev. B* 2008, *77*, 195412.
32. Chen, J. H.; Jang, C.; Adam, S.; Fuhrer, M. S.; Williams, E. D.; Ishigami, M. Charged-Impurity Scattering in Graphene. *Nat. Phys.* 2008, *4*, 377–381.
33. Farmer, D. B.; Chiu, H.-Y.; Lin, Y.-M.; Jenkins, K. A.; Xia, F.; Avouris, P. Utilization of a Buffered Dielectric to Achieve High Field-Effect Carrier Mobility in Graphene Transistors. *Nano Lett.* 2009, *9*, 4474–4478.
34. Miyazaki, H.; Li, S.; Kanda, A.; Tsukagoshi, K. Resistance Modulation of Multilayer Graphene Controlled by the Gate Electric Field. *Semicond. Sci. Technol.* 2010, *25*, 034008 (8pp).
35. Kim, S.; Nah, J.; Jo, I.; Shahrjerdi, D.; Colombo, L.; Yao, Z.; Tutuc, E.; Banerjee, S. K. Realization of a High Mobility Dual-Gated Graphene Field-Effect Transistor with $Al_2O_3$ Dielectric. *Appl. Phys. Lett.* 2009, *94*, 062107.
36. Meric, I.; Han, M. Y.; Young, A. F.; Ozyilmaz, B.; Kim, P.; Shepard, K. L. Current Saturation in Zero-Bandgap, Top-Gated Graphene Field-Effect Transistors. *Nat. Nanotechnol.* 2008, *3*, 654–659.
37. Frank, D.; Taur, Y.; Wong, H.-S. Generalized Scale Length for Two-Dimensional Effects in MOS-FETs. *Electron Device Letters, IEEE* 1998, *19*, 385 –387.
38. Elias, D. C.; Nair, R. R.; Mohiuddin, T. M. G.; Morozov, S. V.; Blake, P.; Halsall, M. P.; Ferrari, A. C.; Boukhvalov, D. W.; Katsnelson, M. I.; Geim, A. K. *et al*. Control of Graphene's Properties by Reversible Hydrogenation: Evidence for Graphane. *Science* 2009, *323*, 610–613.
39. Balog, R.; Jorgensen, B.; Nilsson, L.; Andersen, M.; Rienks, E.; Bianchi, M.; Fanetti, M.; Laegsgaard, E.; Baraldi, A.; Lizzit, S. *et al*. Bandgap Opening in Graphene Induced by Patterned Hydrogen Adsorption. *Nat. Mater.* 2010, *9*, 315–319.
40. Bai, J.; Zhong, X.; Jiang, S.; Huang, Y.; Duan, X. Graphene Nanomesh. *Nat. Nanotechnol.* 2010, *5*, 190–194.
41. Kim, M.; Safron, N. S.; Han, E.; Arnold, M. S.; Gopalan, P. Fabrication and Characterization of Large-Area, Semiconducting Nanoperforated Graphene Materials. *Nano Lett.* 2010, *10*, 1125–1131.
42. Miyazaki, H.; Odaka, S.; Sato, T.;Tanaka, S.; Goto, H.; Kanda, A.; Tsukagoshi, K.; Ootuka, Y.; Aoyagi, Y. Inter-Layer Screening Length to Electric Field in Thin Graphite Film. *Appl. Phys. Express* 2008, *1*, 034007.

FIGURE CAPTIONS

Figure 1. (a) Cross-sectional diagram of the dual-gate graphene FET. Typical optical microscopy images of a device (b) before and (c) after electrode formation. Five FETs are defined on a single graphene flake. (d) Schematic band structures for the BLG before and after application of a perpendicular electric field $E_z$. (e)–(h) $R$ characteristics for both the SLG and BLG FETs at ambient and liquid-nitrogen temperatures. A gating hysteresis arises at RT as shown in (g), but it is almost eliminated at the liquid-nitrogen temperature, as shown in (h).

Figure 2. Output characteristics for both the (a) metallic SLG and (b) semiconducting BLG FETs at $V_{BG}$= -20 V. The left and right longitudinal axes are for the p- and n-type branches, respectively. Two solid dots indicate the bias condition of $V_{TG} - V_{TG}^{0}$ = -0.7 V and $V_{DS}$= -1 V.

Figure 3. Operating principle and electrical characteristics for complementary-like graphene inverters. The applied $V_{BG}$ is 0 V and -30 V for the SLG and BLG inverters, respectively. In the BLG device, large $V_{BG}$ is used to create a large EG (~100 meV) and improve device performance. (a) Wiring schematic. (b) Formation of CNP splitting and (c) resulting voltage inversion under $V_{DD}$ bias in a BLG inverter. (d)–(l) Output response, voltage gain, and normalized channel current as a function of input voltage for both the metallic SLG and semiconducting BLG devices. The $V_{DD}$ is changed from 0.5 to 2 V in (d)–(i), and from 0.2 to 2 V for the BLG device in (j)–(l).

Figure 4. Logic operations for a BLG inverter at 0.1–10 kHz with $V_{DD}$= 1 V. The applied $V_{BG}$ is -6 V to tune the complementary range and achieve a voltage match between input and output. The Boolean "0" and "1" are defined at 0.3 and 0.8 V, respectively.

FIGURES.

Figure 1

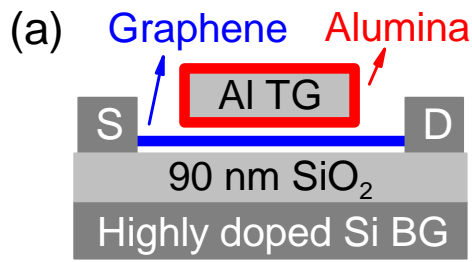

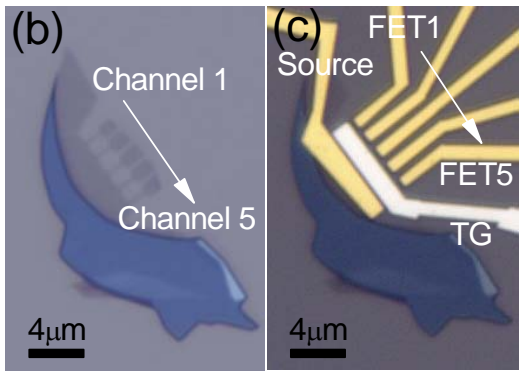

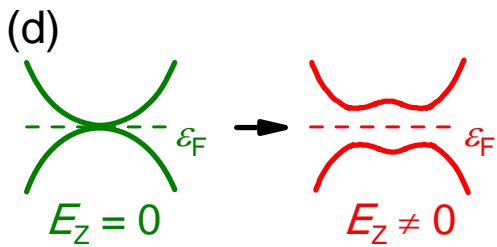

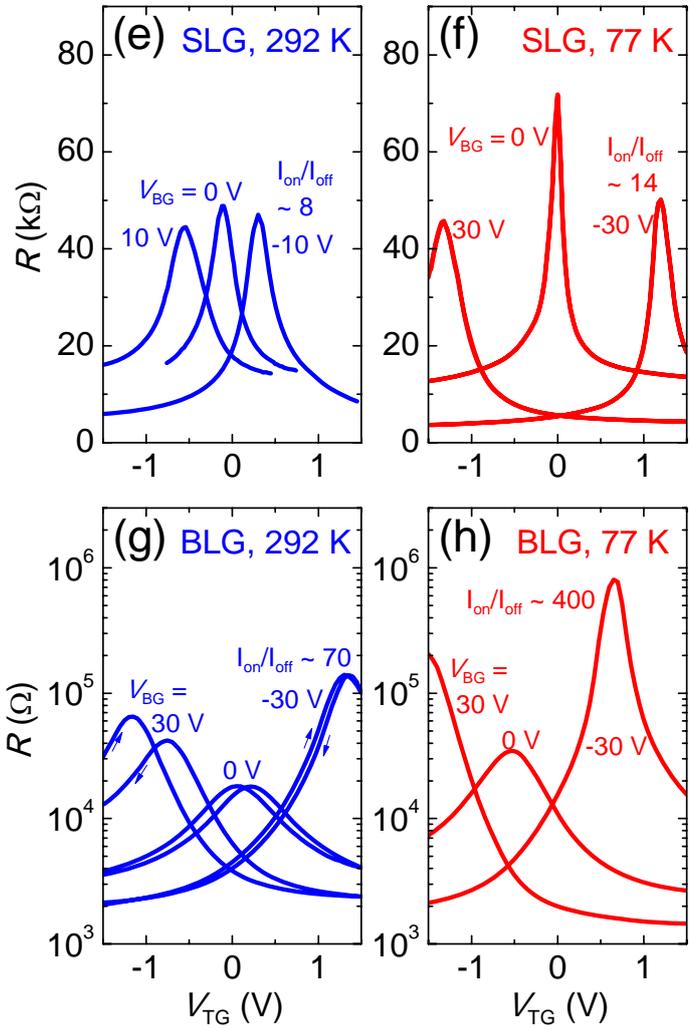

Figure 2

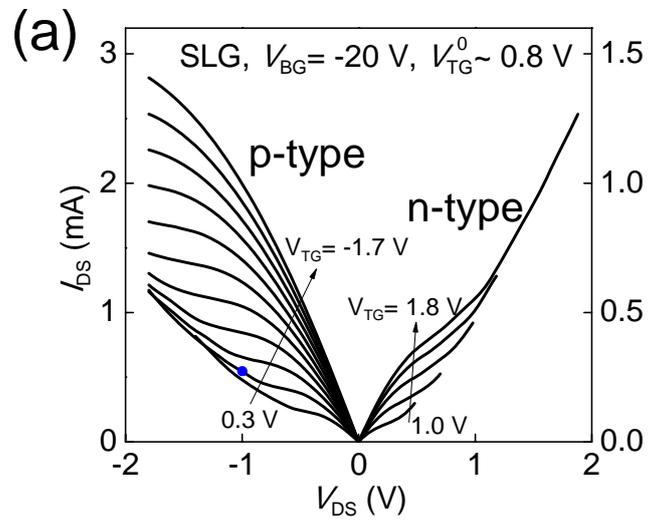

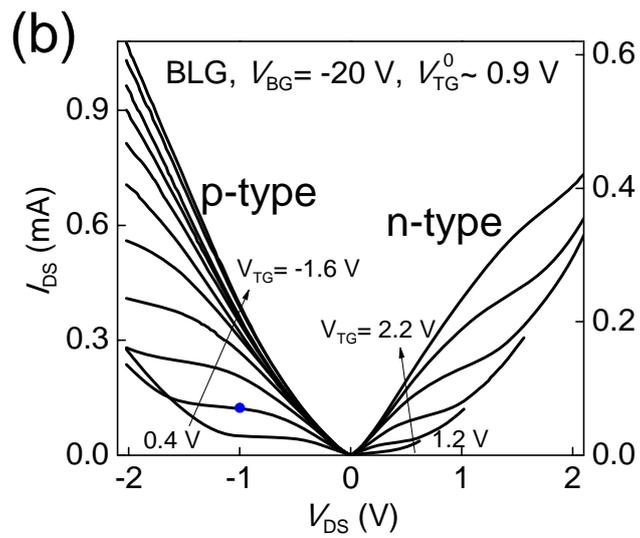

Figure 3

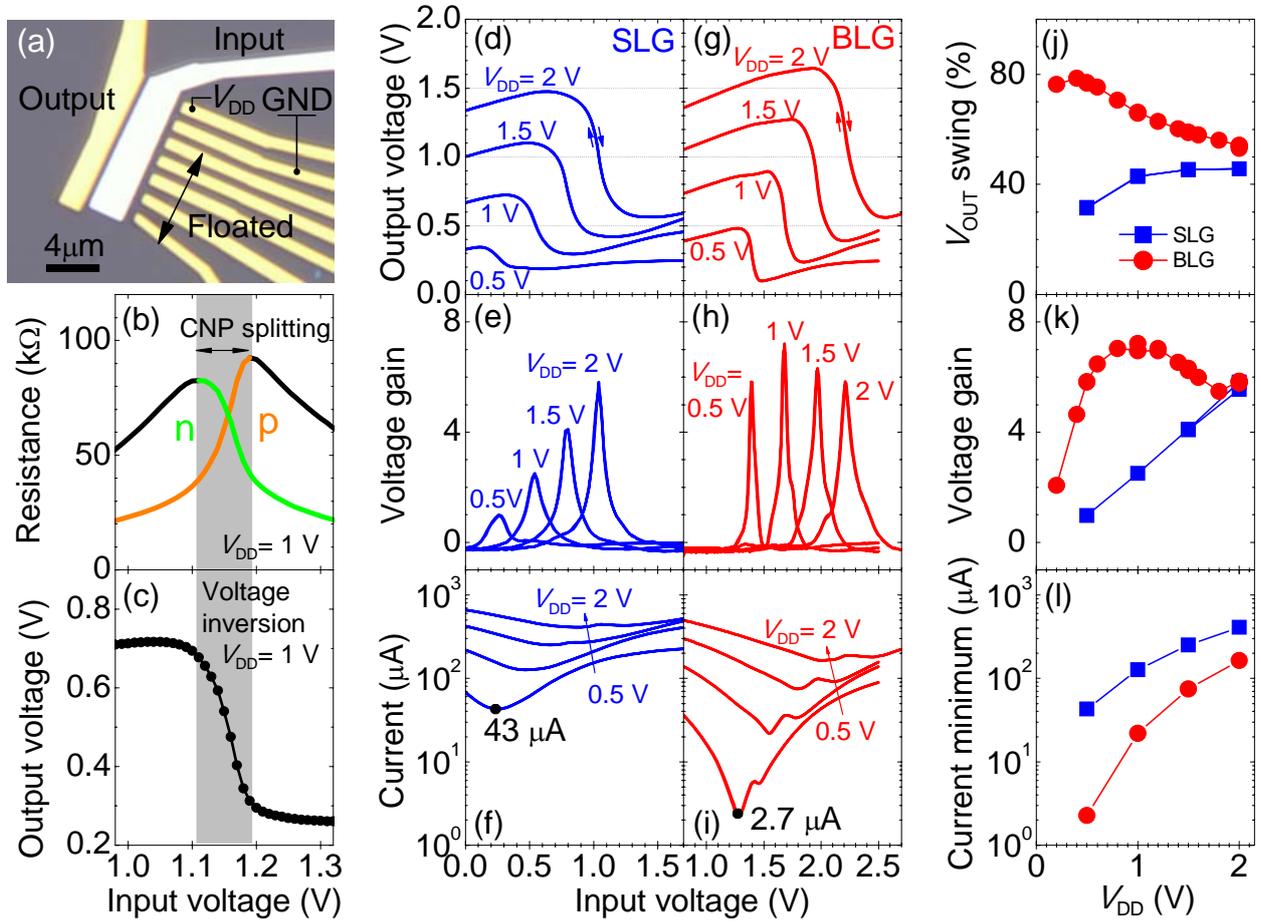

Figure 4

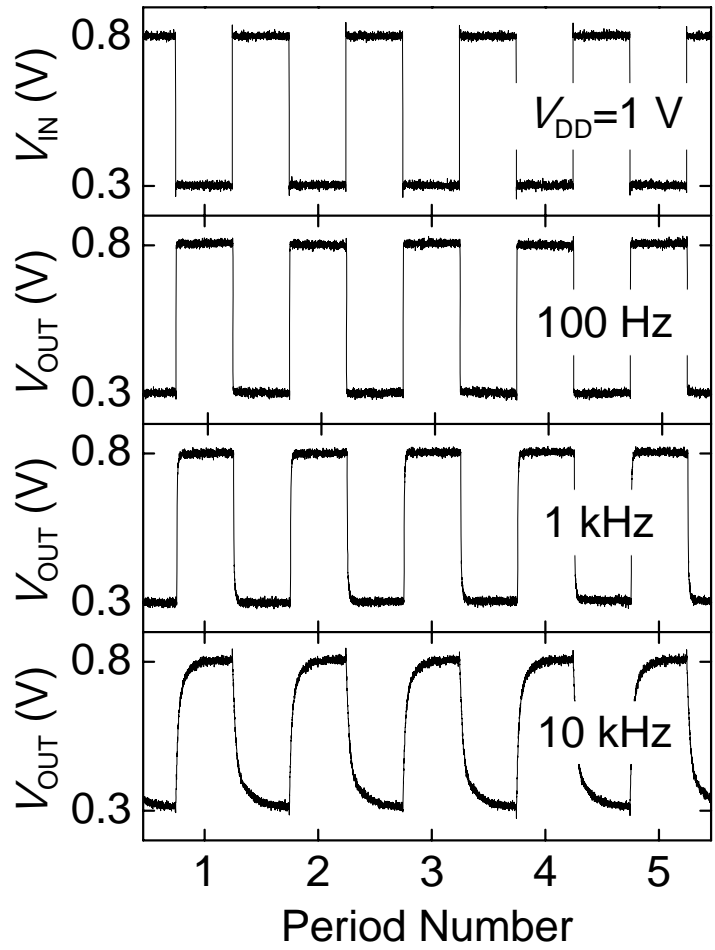

SUPPORTING INFORMATION

**Enhanced Logic Performance with Semiconducting Bilayer Graphene Channels**

Song-Lin Li [1], Hisao Miyazaki [1,2], Hidefumi Hiura [1,3], Chuan Liu [1] and Kazuhito Tsukagoshi [1,2]
[1] MANA, National Institute for Materials Science, Tsukuba, Ibaraki 305-0044, Japan
[2] CREST, Japan Science and Technology Agency, Kawaguchi, Saitama 332-0012, Japan
[3] Green Innovation Research laboratories, NEC Corporation, Tsukuba, Ibaraki 305-8501, Japan

**Content**
1. Reliability of top gate
2. Resistance characteristics for SLG and BLG FETs
3. Output current characteristics for SLG and BLG FETs
4. Estimation of $n_0$ and $\mu$ under buffer-free alumina dielectric
5. Characteristics for BLG inverters under ambient environment
6. Derivation of qualitative expressions for $V_{OUT}$ swing and voltage gain

1. Reliability of top gate

The leakage current and breakdown voltage show no obvious variation between room and low temperatures. As shown in figure S1, the leakage remains below 0.2 nA at 297 K when $V_{TG}$=5V, a negligible value as compared with the µA-order channel current. The breakdown voltage is highly dependent on the device biasing condition. In the extreme case when devices are highly loaded at $V_{BG}$=-30 V and $V_{DS}$=2 V, the breakdown voltage is reduced around 2.4–2.8 V, dependent on devices. Nevertheless, a bias rang within 1.8 V (in the normal case) is very reliable for the top gate at the concerned room temperatures.

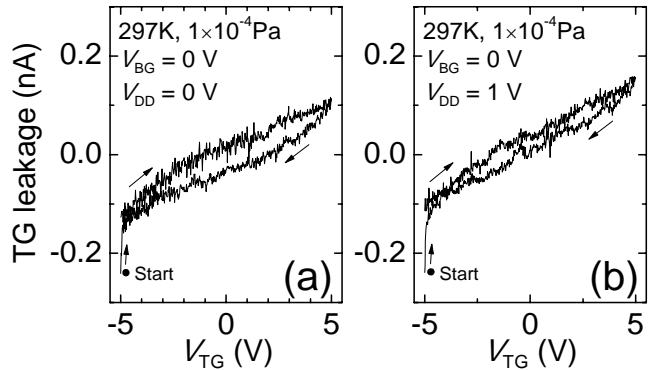

Figure S1. Top gate leakage at room temperature when $V_{DD}$ is biased at (a) 0 V and (b) 1 V.

2. Resistance characteristics for SLG and BLG FETs

As shown in figure S2, the formation of energy gap in the BLG FET can be clearly seen when comparing the resistance response to perpendicular electric fields for the two types of FETs. In SLG, the $R$ peak ($R_{CNP}$) does not essentially change with $V_{BG}$, indicating no change of band structure. The slight decrease with $|V_{BG}|$ stems from the decrease of series resistance $R_s$ (due to the TG-ungated graphene regions around source and drain) as $|V_{BG}|$ increases. In contrast, the $R_{CNP}$ in BLG forms a saddle-like trace, which exhibits a drastic increase with $|V_{BG}|$ (more strictly, perpendicular displacement field $D_z$). At $V_{BG}$= -30 V the $R_{CNP}$ reaches 0.8 MΩ, showing a 70 times enhancement to the case of $V_{BG}$= 0 V. Such a tendency is consistent with the theoretical prediction on opening and expansion of EG with $D_z$.

We can also estimate the TG capacitive efficiency from the $R_{CNP}$ trace in the 2D resistance plot, in which a linear relation between $V_{TG}$ and $V_{BG}$ can be derived. When $V_{BG}$ is varied from -30 to 30 V, the CNP position changes accordingly from ~1.2 to ~-1.3 V at the $V_{TG}$ axis. This indicates that the 2.5 V change of $V_{TG}$ is equivalent to 60 V of $V_{BG}$ (with 90 nm SiO$_2$ dielectric). Thus, the coupling capacitance of TG ($C_{TG}$) is 24 times of that of BG, i.e., $C_{TG}/C_{BG} = 24$. Using the geometry capacitance of BG, the $C_{TG}$ is estimated to be ~920 nF/cm$^2$, representing one of the thinnest and highest efficiency oxide dielectrics used in graphene FETs.[S1,S2] We also emphasize that our TG stack technique is remarkably simple which simultaneously defines the TG metal and dielectric layers by a one-time aluminum deposition. The dielectric layer accomplishes naturally in the sequent passivation process when devices are exposed to air.[S1,S3]

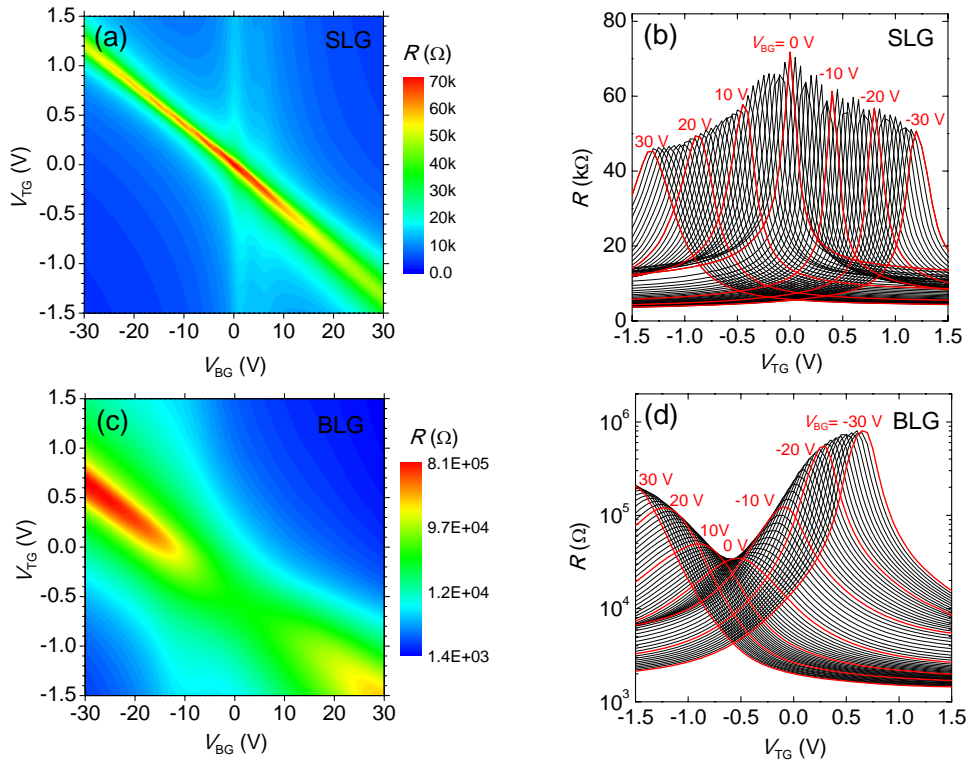

Figure S2. Comparison of the resistance behavior for individual SLG and BLG FETs at 77 K. (a) and (c) show the two-dimensional resistance plot as a function of $V_{BG}$ and $V_{TG}$. Note that the $R$ is plotted in a linear scale in (a) and a logarithmic scale in (c). (b) and (d) show the corresponding plots as $V_{TG}$ sweeps.

3. Output current characteristics for SLG and BLG FETs

Figure S3 presents a full contour plot of $I_{DS}$ as a function of $V_{DS}$ and $V_{TG}$ for both the SLG and BLG FETs. For the gapless SLG FET, the unique second linear region is observed (Fig. S3a), which is due to the high capacitive efficiency and consistent with the report of Ref. S4. The current saturation behavior becomes improved when small energy gap is introduced in the channel in the BLG FET, as reflected by the stronger current "kink" in Fig. S3b.

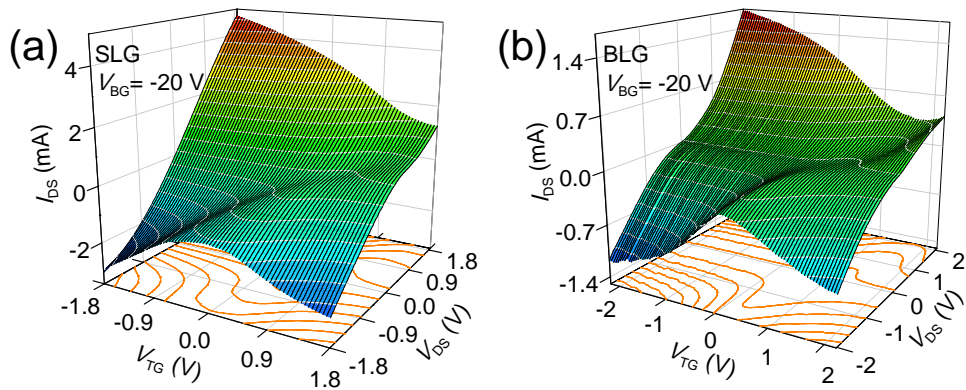

Figure S3. Full contour plots of $I_{DS}$ behavior for individual SLG and BLG FETs at 77 K.

4. Estimation of $n_0$ and $\mu$ under buffer-free alumina dielectric

The TG bias $V_{TG}$ not only creates an electrostatic potential difference $\phi$ between TG and graphene, but also shifts the Fermi level ($E_F$) of graphene due to the change of carrier density $n$. Therefore (Ref. S5),

$$V_{TG} = \phi + \frac{E_F}{e} \quad (1)$$

where the first term $\phi = en/C_{TG}$ is related to TG dielectric and the second term $E_F/e$ is related to quantum capacitance $C_Q$.

To accurately estimate the residual carrier density $n_0$ and mobility $\mu$ of graphene FETs, the effect of $R_s$, contact resistance $R_c$, and quantum capacitance $C_Q$ should be considered. As suggested by Ref. R6, the carrier density and resistance for SLG ($E \approx v_F p$) follow the relations below:

$$\Delta V = |V_{TG} - V_{TG,CNP}| = \frac{en}{C_{TG}} + \frac{\hbar v_F \sqrt{\pi n}}{e} \quad (2)$$

$$R_{tot} = R_c + R_s + \frac{L}{We\mu\sqrt{n_0^2 + n^2}} \quad (3)$$

where $v_F = 1 \times 10^6$ m/s is the Fermi velocity, $e$ is the elementary charge, $R_{tot}$ is the total resistance, $L$ is the channel length, $W$ is the channel width and $n_0$ is the residual carrier density at Dirac point. Corresponding fits at $T$=292 and 77 k are shown in Figs. S4a and S4b.

The band structure of BLG is different to that of SLG. In experiment related intermediate energy range, the two low-energy bands can be approximated as [S7]

$$E^{\pm} = \pm \tfrac{1}{2}\gamma_1(\sqrt{1 + 4v_F^2 p^2/\gamma_1^2} - 1) \quad (4)$$

It evolves as a quadratic spectrum $E \approx v_F^2 p^2/\gamma_1$ at low $p$, and as a linear spectrum $E \approx v_F p$ as SLG at high $p$. The crossover happens at $n \approx 4.4 \times 10^{12}$ cm$^{-2}$ which corresponds to $V_{TG} \approx 0.7$ V in our devices. In two-dimensional systems, $p^2 = \hbar^2 k^2 = \hbar^2 \pi n$. For accurate fit of BLG, one should use the $E_F$–$n$ relation extracted from Eq. 4 and employ

$$\Delta V = \frac{en}{C_{TG}} + \frac{\gamma_1\left(\sqrt{1+(\hbar v_F)^2 \pi n} - 1\right)}{2e} \quad (5)$$

to replace Eq. (2). However, we found that rather reasonable results can still be obtained even adopting the fitting equations for SLG, as shown in Figs. S4c and S4d. All the fitting results for SLG and BLG are shown in Table S1.

Table S1. Fitted mobility $\mu$ and residual carrier density $n_0$ for SLG and BLG channels under buffer-free alumina dielectric in the ambient (air, 292 K) and liquid nitrogen (N$_2$, 77 K) environments.

|  | SLG (air, 292 K) | SLG (N$_2$, 77 K) | BLG (air, 292 K) | BLG (N$_2$, 77 K) |
|---|---|---|---|---|
| $\mu$ (cm$^2$/Vs) | 5000 | 8500 | 1600 | 1100 |
| $n_0$ (10$^{12}$/cm$^2$) | 0.35 | 0.14 | 1.1 | 0.084 |

The slight reduction of $n_0$ for SLG in liquid nitrogen surrounding is attributed to the reduced charged gas adsorbates. In addition, the large reduction of $n_0$ and small decrease of $\mu$ in BLG in Fig. S4d are likely due to the presence of a small band gap, since the BLG is slightly biased ($V_{TG,CNP}$=-0.55 V and $V_{BG}$=0).

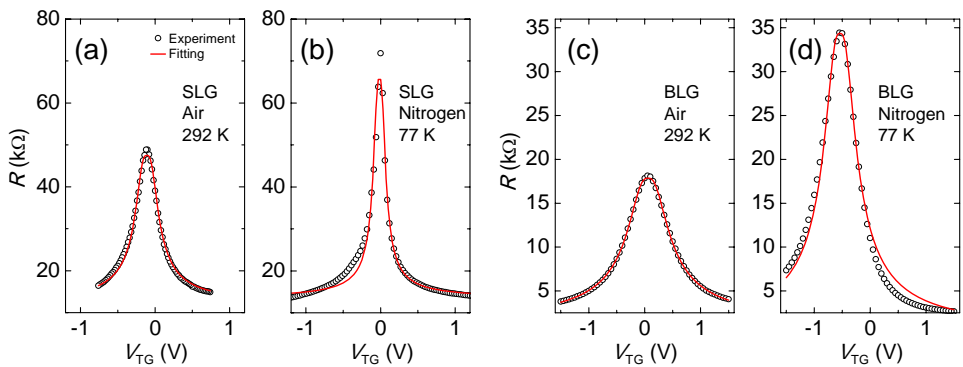

Figure S4. Mobility $\mu$ and residual carrier density $n_0$ fittings with quantum capacitance included. (a) SLG under ambient condition (air, 292 K), (b) SLG in liquid nitrogen (N$_2$, 77 K), (c) BLG under ambient condition (air, 292 K), (b) BLG in liquid nitrogen (N$_2$, 77 K).

5. Characteristics for BLG inverters under ambient environment

The hysteresis in ambient is mainly caused by two opposite dynamic processes: a fast gas desorption at high current and a slightly slower re-absorption at low current in air, because it almost disappears in vacuum at room temperature. Usually, it is rather large (~ 0.2 V) for samples kept for long time, and becomes smaller and smaller with constant sweeps (anneal by Joule heat). Figure S5 shows the voltage transfer characteristic and corresponding voltage gain for a BLG inverter in ambient environment. The hysteresis is largely reduced after constant annealing. This may indicate a balance between the desorption and absorption processes.

As indicated in the manuscript, a decrease of device performance is expected due to the limited energy gap in our BLG devices and enhanced thermal energy for charge carriers at room temperature. However, the $V_{OUT}$ swing still reaches 50% (Fig.S5a) and the voltage gain remains larger than 1, increasing from 1.5 to 1.9, as $V_{BG}$ changes from -18 to -20 V (Fig.S5b).

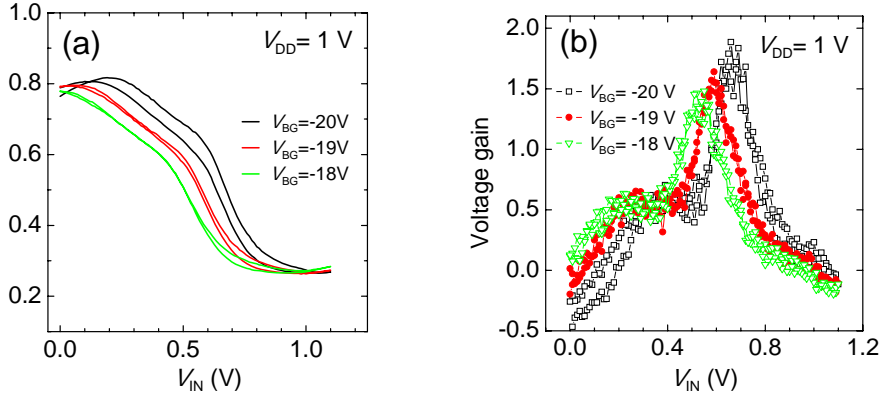

Figure S5. Typical voltage transfer characteristic and corresponding voltage gain for a BLG inverter in ambient environment ambient environment.

6. Derivation of qualitative expressions for $V_{OUT}$ swing and voltage gain

Accurate analytical expressions for the voltage gain and $V_{OUT}$ swing should be rather complex and beyond the scope of this study. For simplicity, qualitative expressions are devived by neglecting $R_s$ and $R_c$ and assuming a linear $\sigma - V_{TG}$ relation. Under such assumptions, the conductance of the BLG channel behaves as two symmetric straight lines with respect to the charge neutrality point (CNP) and can be written as

$$\sigma = \sigma_0 + k\left|V_{TG} - V_{TG}^0\right|$$

where $k$ is the tunability of conductance through gate voltage, $\sigma_0$ is the off-state (CNP) conductance, and $\Delta$ is the CNP splitting between the two FETs along $V_{IN}$ axis. In real FETs, a rather smooth (parabolic-like) shape of the curve is expected near CNP regions; the employed sharp $\sigma$ transition is also for simplifying the derivations. We note that the simplifications involved above still warrant the correctness of the final results in a qualitative way. We also assume that the $\sigma_0 - V_{TG}$ curve shapes of the two involved FETs are identical, except for different $V_{TG}^0$.

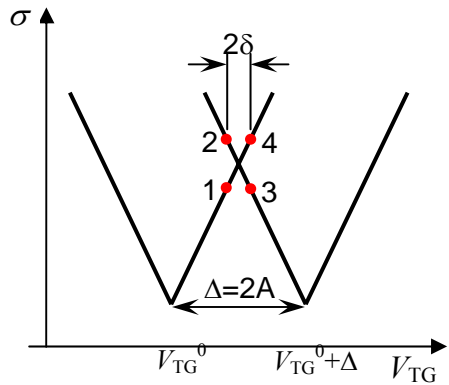

Figure S6. Schematic conductivity curves for the two involved graphene FETs in a complementary-like inverter. The splitting between the neutrality points is induced by $V_{DD}$. The voltage inversion behavior is formed in the splitting range.

For convenience, we use CNP splitting $\Delta = 2A$ and the distance between points is $2\delta$ around the cross point. The two characteristic $\sigma$ values are

$$\sigma_1 = \sigma_3 = \sigma_0 + k(A - \delta),$$
$$\sigma_2 = \sigma_4 = \sigma_0 + k(A + \delta).$$

Correspondingly, the characteristic $R$ values are

$$R_1 = R_3 = \frac{1}{\sigma_0 + k(A - \delta)},$$
$$R_2 = R_4 = \frac{1}{\sigma_0 + k(A + \delta)}.$$

Use $V_{12}$ and $V_{43}$ to represent the output voltages at $V_{TG}=V_0+A-\delta$ and $V_{TG}=V_0+A+\delta$, respectively. We obtain

$$\Delta V = V_{12} - V_{43} = \frac{R_1}{R_1 + R_2}V_{DD} - \frac{R_4}{R_4 + R_3}V_{DD} = \frac{R_1 - R_2}{R_1 + R_2}V_{DD} = \frac{k\delta}{\sigma_0 + kA}V_{DD}.$$

Then, the qualitative expressions of $V_{OUT}$ swing and voltage gain are

$$V_{OUT} \text{ Swing } = \Delta V / V_{DD}|_{\delta = A} = \frac{1}{1 + 2\sigma_0/k\Delta}$$
$$\text{Voltage Gain} = \Delta V / 2\delta|_{\delta \to 0} = \frac{V_{DD}}{2\sigma_0/k + \Delta}.$$

**Reference**

(S1) Miyazaki, H.; Li, S.; Kanda, A. & Tsukagoshi, K. (2010), 'Resistance Modulation of Multilayer Graphene Controlled by the Gate Electric Field', *Semicond. Sci. Technol.* **25**(3), 034008.
(S2) Wang, Z.; Xu, H.; Zhang, Z.; Wang, S.; Ding, L.; Zeng, Q.; Yang, L.; Pei, T.; Liang, X.; Gao, M. & Peng, L.-M. (2010), 'Growth and Performance of Yttrium Oxide as an Ideal High-κ Gate Dielectric for Carbon-Based Electronics', *Nano Lett.* **10**(6), 2024--2030.
(S3) Miyazaki, H.; Odaka, S.; Sato, T.; Tanaka, S.; Goto, H.; Kanda, A.; Tsukagoshi, K.; Ootuka, Y. & Aoyagi, Y. (2008), 'Inter-Layer Screening Length to Electric Field in Thin Graphite Film', *Appl. Phys. Express* **1**(3), 034007.
(S4) Meric, I.; Han, M. Y.; Young, A. F.; Ozyilmaz, B.; Kim, P. & Shepard, K. L. (2008), 'Current Saturation in Zero-Bandgap, Top-Gated Graphene Field-Effect Transistors', *Nat. Nanotechnol.* **3**(11), 654-659.
(S5) Das, A.; Pisana, S.; Chakraborty, B.; Piscanec, S.; Saha, S. K.; Waghmare, U. V.; Novoselov, K. S.; Krishnamurthy, H. R.; Geim, A. K.; Ferrari, A. C. & Sood, A. K. (2008), 'Monitoring dopants by Raman Scattering in an Electrochemically Top-Gated Graphene Transistor', *Nat. Nanotechnol.* **3**(4), 210-215.
(S6) Kim, S.; Nah, J.; Jo, I.; Shahrjerdi, D.; Colombo, L.; Yao, Z.; Tutuc, E. & Banerjee, S. K. (2009), 'Realization of a High Mobility Dual-Gated Graphene Field-Effect Transistor with Al2O3 DIELECTRIC', *Appl. Phys. Lett.* **94**(6), 062107.
(S7) McCann, E. & Fal'ko, V. (2006), 'Landau-Level Degeneracy and Quantum Hall Effect in a Graphite Bilayer', *Phys. Rev. Lett.* **96**(8), 086805.